\documentclass[final, nomarks
]{dmtcs-episciences}

\usepackage[utf8]{inputenc}
\usepackage{subfigure}

\usepackage[round]{natbib}

\author{Lucas Meijer \thanks{Lucas Meijer is generously supported by the Netherlands Organisation for Scientific Research (NWO) under project no. VI.Vidi.213.150.} \and Tillmann Miltzow \thanks{Tillmann Miltzow is generously supported by the Netherlands Organisation for Scientific Research (NWO) under project no. 016.Veni.192.250 and no. VI.Vidi.213.150.}}

\title{Sometimes Two Irrational Guards are Needed}

\affiliation{
  Utrecht University, Utrecht, The Netherlands}

\usepackage[usenames,dvipsnames]{xcolor}
\usepackage[T1]{fontenc}
\usepackage[english]{babel}
\usepackage[utf8]{inputenc}
\usepackage{amsmath,amssymb,amsfonts,mathrsfs,amsthm}
\usepackage{mathtools}
\usepackage{stmaryrd}
\usepackage{thm-restate}
\usepackage[capitalise]{cleveref}
\usepackage{xspace}
\usepackage{float}
\usepackage{comment}

\numberwithin{equation}{section}

\newtheorem{theorem}{Theorem}

\theoremstyle{definition}

\crefname{figure}{Figure}{Figures}
\Crefname{figure}{Figure}{Figures}
\crefname{claim}{Claim}{Claims}
\Crefname{claim}{Claim}{Claims}
\crefname{observation}{Observation}{Observations}
\Crefname{observation}{Observation}{Observations}

\newcommand{\Q}{\ensuremath{\mathbb{Q}}\xspace}

\newcommand{\R}{\ensuremath{\mathbb{R}}\xspace}

\renewcommand{\epsilon}{\ensuremath\varepsilon}

\renewcommand{\phi}{\ensuremath{\varphi}}

\renewcommand{\epsilon}{\ensuremath{\varepsilon}}

\renewcommand{\theta}{\ensuremath{\vartheta}}

\newcommand{\ER}{\ensuremath{\exists\mathbb{R}}\xspace}
\newcommand{\NP}{\text{NP}\xspace}
\newcommand{\PSPACE}{\text{PSPACE}\xspace}
\newcommand{\problemname}[1]{\textnormal{\textsc{#1}}\xspace}
\newcommand{\ETR}{\problemname{ETR}}

\newcommand{\wordRAM}{\textnormal{word RAM}\xspace}
\newcommand{\realRAM}{\textnormal{real RAM}\xspace}

\newcommand{\frontRAY}{\textnormal{front ray}\xspace}
\newcommand{\backRAY}{\textnormal{back ray}\xspace}
\newcommand{\feasibleSegment}{\textnormal{feasible segment}\xspace}

\newcommand{\vis}{\ensuremath{\text{vis}}\xspace}

\newcommand{\GuardLucas}{\ensuremath{l}\xspace}
\newcommand{\GuardTill}{\ensuremath{t}\xspace}

\begin{document}

\publicationdata{vol. 28:2}{2026}{10}{10.46298/dmtcs.11563}{2023-07-10; 2023-07-10; 2024-06-21}{2026-01-21}

\maketitle

\begin{abstract}
    In the art gallery problem, we are given a closed polygon $P$, with rational coordinates and 
    an integer~$k$. 
    We are asked whether it is possible to find a set (of guards) $G$ of size $k$
    such that any point $p\in P$ is seen by a point in $G$.
    We say two points $p$, $q$ see each other if the line segment $pq$ 
    is contained inside $P$.
    It was shown by Abrahamsen, Adamaszek, and Miltzow that there is a polygon
    that can be guarded with three guards, but requires four guards if the
    guards are required to have rational coordinates.
    In other words, an optimal solution of size three might need to be irrational.
    We show that an optimal solution of size two might need to be irrational.
    Note that it is well-known that any polygon that can be guarded with one guard
    has an optimal guard placement with rational coordinates.

    Hence, our work closes the gap on when irrational guards are possible to occur.
\end{abstract}

\date{}
%%%%%%%%%%%%%%%%%%%%%%%%%%%%%%%%%%%%%%%%%%%%%%%%%%%%%%%%%%%%%%

\maketitle

\vspace{1cm}

\begin{figure}[ht]
    \centering
    \includegraphics{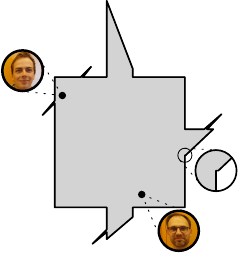}
    \caption{Lucas and Till guarding the polygon with just the two of them...}
    \label{fig:title}
\end{figure}

\newpage
\section{Introduction}
    In the art gallery problem, we are given a closed polygon $P$, on 
    $n$ vertices, with rational coordinates and 
    an integer $k$.
    We are asked whether it is possible to find a set (of guards) $G$ of size $k$
    such that any point $p\in P$ is seen by a point in $G$.
    We say two points $p$, $q$ see each other if the line segment $pq$ 
    is contained inside $P$.
    We show that an optimal solution of two guards might need to have irrational coordinates.
    In such a case, we say a polygon has \emph{irrational guards}.

    The art gallery problem was formulated in 1973 by Victor Klee. See, for example, the book by O'Rourke \cite[page 2]{o1987art}.
    One of the earliest results states that every simple polygon on $n$ vertices can always be guarded with $\lfloor n/3 \rfloor$ guards~\cite{chvatal1975combinatorial,Fisk78a}.

    \begin{figure}[bp]
        \centering
        \includegraphics{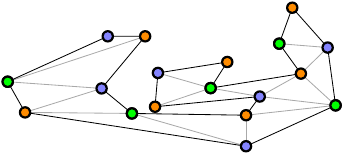}
        \caption{Any triangulation of a simple polygon can be three-colored.
        At least one of the color classes has at most $\lfloor n/3\rfloor$ vertices. This color class also guards the entire polygon, as every triangle is incident to all three colors~\cite{Fisk78a}.}
        \label{fig:Fisk}
    \end{figure}

    Interestingly, it is actually very tough to find any positive algorithmic results on the art gallery problem. It seems like the art gallery problem is almost impenetrable.
    For instance, only in 2002, Micha Sharir pointed out that the problem was even decidable~\cite[see acknowledgments]{EfratH02,EfratH06}.
    The decidability of the art gallery problem is actually easy once you know methods from real algebraic geometry~\cite{Basu2006_RealAlgebraicGeometry}.
    The idea is to reduce the problem to the first-order theory of the reals. 
    We encode guard positions by variables, and then we check if every point in the polygon is seen by at least one guard.
    Note that this is easy to encode in the first-order theory of the reals, as we are allowed to use existential ($\exists g_1,g_2,\ldots$) and universal quantifiers ($\forall p =(x,y)$).
    Since then, despite much research on the art gallery problem, no better algorithm appeared, as far as worst-case complexity is concerned. 
    The underlying reason for the difficulty to find better algorithms 
    can be explained by the fact that the art gallery problem is \ER-complete~\cite{stade2022complexity, Abrahamsen2022_artGallery}.
    In a nutshell, \ER-completeness precisely entails that there is no better method for the worst-case complexity of the problem. 
    (\ER can be defined as the class of problems that are equivalent to finding a real root to a multivariate polynomial with integer coordinates. See \Cref{sub:ETR} for an introduction.)
    More specifically, it was shown that arbitrary algebraic numbers may be needed to describe an optimal solution to the art gallery problem.
    This may come as a surprise to some readers, and was clearly a surprise back then.
    Specifically, ``in practice'', it seems very rare that irrational guards are ever needed. 
    The reason is that a typical situation is one of the following two. 
    Either the guards have some freedom to move around and still see the entire polygon.
    Or if a guard has no freedom, it is forced to be on a line defined by vertices of the polygon.
    As the vertices of the polygon are at rational coordinates, the guards will be at rational coordinates in that case as well.
    Indeed, only in 2017, the first polygon requiring irrational guards was found~\cite{abrahamsen2017irrational}.
    Even though \ER-reductions exhibit an infinite number of polygons that require irrational guards, those polygons are not ``concrete'' in the naive sense of the word. 
    And up to this day, this is the only ``concrete'' polygon~\cite{abrahamsen2017irrational} that we know does require irrational guards.
    In this work, we find a second polygon. 
    It is superior to the first one in the sense that it shows that
    two guards are already enough to enforce irrational guards.
    As a single guard can always be chosen to have rational coordinates,
    we settle the question of the minimum number of guards required to
    have irrational guards.

    The polygon we find is again monotone.
    A polygon is called monotone if there exists a line $l$ such that every line orthogonal to $l$ intersects $P$ at most twice.
    We summarize our results in the following theorem.

\begin{theorem}\label{thm:main}
    There exists a simple monotone polygon with rational coordinates, such that there is only one way of guarding this polygon optimally with two guards.
    Those two guards have irrational coordinates.
\end{theorem}

%%%%%%%%%%%%%%%%%%%%%%%%%%%%%%%%%%%%%%%%%%%%%%%%%%%%%%%%%%%%%
\paragraph{Organization.}
    We first discuss our results from different angles (\Cref{sub:discussion}).
    Then we give a selected overview of related research on the art gallery problem (\Cref{sub:art}).
    We finish this introduction with some background on the existential theory of the reals (\Cref{sub:ETR}).
    In \Cref{sec:preparation}, we give an overview of how we constructed the polygon and what is the intuition behind the different parts.
    In \Cref{sec:Polygon}, we give the polygon with coordinates of all vertices and we provide a formal proof of correctness.
    In \Cref{sec:Challenges}, we explain how we constructed the polygon and what technical challenges we had to overcome.

%%%%%%%%%%%%%%%%%%%%%%%%%%%%%%%%%%%%%%%%%%%%%%%%%%%%%%%%%%%%%
\subsection{Discussion}
\label{sub:discussion}

In this section, we discuss different aspects of our findings.

\paragraph{Minimum number of irrational guards.}
It is known that one guard can always be chosen to be rational~\cite{LeePreparataoptimal}.
The polygon by Abrahamsen, Adamaszek, and Miltzow~\cite{abrahamsen2017irrational} requires three irrational guards.
The main strength of our finding is to determine the minimum number of guards required to have irrational guards.

\paragraph{Grid Approximation.}
One way to circumvent worst-case complexity is to discretize the polygon and restrict oneself to a dense grid~\cite{BonnetM17Approx,EfratH02,EfratH06}. 
The polygon by Abrahamsen, Adamaszek, and Miltzow showed that a grid cannot have a better approximation factor than $4/3 = 1.333\ldots$.
We improve this lower bound to $3/2 = 1.5$.

\begin{figure}[bp]
    \centering
    \includegraphics[]{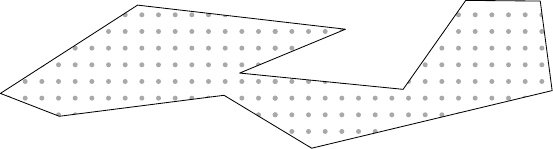}  
    \caption{We may restrict the guards to lie on a dense grid.
    This may make the optimal solution worse.}
    \label{fig:Grid-Approx}
\end{figure}

Note that Bonnet and Miltzow showed that under some mild assumptions, the grid contains a constant factor approximation~\cite{BonnetM17Approx}.

\paragraph{Intuitive Understanding.}
It is good to have multiple different concrete polygons that require irrational guards. 
This result complements the \ER-completeness of the art gallery problem nicely.
While \ER-completeness is clearly stronger from a theoretical perspective, concrete polygons may be useful to get a better intuitive understanding of the difficulty.

\paragraph{Test Cases.}
From a practical perspective, our polygon can serve as a test case on which we can compare the performance of different algorithms.
Usually, we would like to have a host of difficult and diverse instances
that can be automatically generated.
With difficult instances, we mean polygons that require irrational guards.
We leave this as a future research question.

\paragraph{Technical Depth.}
The principal methods that we used for the construction of our polygon are in spirit similar to the methods used by Abrahamsen, Adamaszek, and Miltzow~\cite{abrahamsen2017irrational}.
However, it turned out that it was considerably more difficult to find the polygon.
On the one hand, our construction is smaller and thus there were fewer parameters that we had to manipulate to find a solution.
On the other hand, the two guards interact in more intertwined ways.
Thus making it much harder to find a correct placement of all the polygon vertices.

To be concrete, the construction by \cite{abrahamsen2017irrational} has the guards $a$, $m$, and $t$.
Guards $a$ and $m$ cover together two pockets.
Similarly, guards $m$ and $t$ cover together two separate pockets.
Thus there is no direct interaction between guards $a$ and $t$.
This makes it easier to construct the different parts independently.

In our case the two guards \GuardLucas and \GuardTill together guard three pockets.
And this implies that the interaction between \GuardLucas and \GuardTill is much more integrated. 
This leads to a construction with some vertices being extremely close.
Furthermore, our construction has pockets that were inside other pockets.

\paragraph{Irrational Boundary Guards.}
Both our polygon and the polygon by~\cite{abrahamsen2017irrational} have their 
guards in the interior.
It is an interesting open problem if one can enforce irrational guards, in case all guards are restricted to lying on the boundary and are only required to guard the boundary.

\paragraph{Integer Coordinates.}
One may wonder whether there is also a polygon with integer coordinates that exposes irrational guards.
The answer is yes, and this can be achieved by scaling all coordinates by all the appearing denominators.

\paragraph{Special Polygons.}
One may wonder whether it is possible to enforce irrational guards on
polygons with some extra properties like being rectilinear or monotone. 
Both of those questions have been positively resolved by~\cite{abrahamsen2017irrational}.

%%%%%%%%%%%%%%%%%%%%%%%%%%%%%%%%%%%%%%%%%%%%%%%%%%%%%%%%%%%%%
\subsection{Art Gallery Problem}
\label{sub:art}

The literature on the art gallery is vast. 
Therefore, we decide to focus here on algorithmic results.
\paragraph{Exact Algorithmic Results.}
In 1979, the first algorithm for guarding a polygon in linear time with one guard appeared~\cite{LeePreparataoptimal}.
It took until 1992 until there was an algorithm that could determine if a polygon could be guarded by two guards in $O(n^4)$ time~\cite{BellevilleCCCG,belleville1991computing}.
As mentioned already above it took until 2002 to find the first correct algorithm that solves the art gallery problem~\cite{EfratH02, EfratH06}.
There is still no other algorithm known.

On the lower bound side, we know \NP-hardness~\cite{LeeLin86}, APX-hardness~\cite{eidenbenz2001inapproximability} and W[1]-hardness~\cite{BonnetW1HARD}.
One may argue that \NP-hardness is enough evidence that there are
no efficient algorithms for the art gallery problem and that this
may fully explain the lack of algorithmic results.
However, for other \NP-complete problems like Clique, Subset-sum, Dominating Set, and TSP, we do know a myriad of algorithms. 
Although many of them run in exponential time in the worst case they give huge improvements in many different situations.
We believe that the lack of algorithmic results may stem from the fact that
we do not know how to discretize the art gallery problem efficiently.
Note that all the mentioned problems are already discrete.
We believe that the \ER-completeness of the art gallery problem may give the most compelling explanation of why a concise discretization of the art gallery problem is unlikely~\cite{Abrahamsen2022_artGallery,stade2022complexity}.
Specifically, many discretization schemes would imply that the art gallery problem lies in \NP and thus imply $\NP = \ER$.
The first proof that the art gallery problem is \ER-comlete was given by Abrahamsen, Adamaszek, and Miltzow in 2017~\cite{Abrahamsen2022_artGallery}.
It was recently improved by Stade, who showed that it is even \ER-complete if we only require the boundary to be guarded~\cite{stade2022complexity}.
It remains open whether guarding the boundary from the boundary is \ER-complete.

\paragraph{Practical Difficulty.}
There is a series of papers that studied the art gallery problem
from a practical perspective. 
In other words, they implemented algorithms and tested them on benchmark instances~\cite{PracticalBorrmann, PracticalARTMasterFriedrich, Simon-hengeveld2021practical, tozoni, Hengeveld2021_artGallery}.
The practical experiences from those papers suggest that irrational coordinates do not play any role in the pursuit to find an optimal solution.

\begin{figure}[bp]
\centering
    \includegraphics[page = 4]{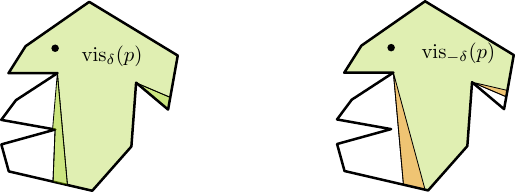}
    \caption{Left: The dark green region is added to the visibility polygon. Right: The orange region is removed from the visibility polygon.}
\label{fig:SmallChanges}
\end{figure}

\paragraph{Explaining the Discrepancy.}
We are aware of two theoretical explanations for the discrepancy between the theoretical and the practical results.
One such finding uses smoothed analysis and argues that there is with high probability an optimal solution after a small random perturbation of the polygon~\cite{Hengeveld2021_artGallery, Erickson2022_SmoothingGap}.
A second explanation comes from Hengeveld and Miltzow.
They introduced the notion of vision-stability.
To explain this concept, we consider guards that can either see by some small angle $\delta$ around reflex vertices or are blocked by an angle $\delta$ by reflex vertices. 
See the green visibility regions in \Cref{fig:SmallChanges}.

Intuitively, if $\delta$ is small enough then the optimal number of guards will not change.
Vision-stability states that there indeed exists such a $\delta>0$.
Using this assumption Hengeveld and Miltzow could find a polynomially-sized discretization scheme for the art gallery problem.
It remains an open question to improve their discretization scheme.

\paragraph{Topology.}
Given a polygon, we can consider the set $\mathcal{G}(P)$ of all possible guard sets of minimum size. 
Together with the Hausdorff distance, $\mathcal{G}(P)$ forms a topological space.
From the algebraic encoding by Sharir, we know that the topological space must be compact and semi-algebraic.
The question arises: given a compact semi-algebraic $S$, is there a polygon $P$ such that $\mathcal{G}(P)$ is topologically equivalent to $S$?

This question was positively answered by Bertschinger, El Maalouly, Miltzow,  Schnider, and Weber for homotopy-equivalence~\cite{TOPOLOGY-SIMON} and shortly improved by Stade and Tucker-Foltz to homeomorphic-equivalence~\cite{stade2022topological}.

%%%%%%%%%%%%%%%%%%%%%%%%%%%%%%%%%%%%%%%%%%%%%%%%%%%%%%%%%%%%%
\subsection{Existential theory of the Reals}
\label{sub:ETR}

The complexity class \ER (often pronounced as ``ER'') has gained a lot of interest in recent years.
It is defined via its canonical complete problem \ETR (short for \emph{Existential Theory of the Reals}) and contains all problems that polynomial-time many-one reduce to it.
In an \ETR instance, we are given an integer~$n$ and a sentence of the form
\[
    \exists X_1, \ldots, X_n \in \R :
    \varphi(X_1, \ldots, X_n),
\]
where~$\varphi$ is a well-formed and quantifier-free formula consisting of polynomial equations and inequalities in the variables and the logical connectives $\{\land, \lor, \lnot\}$.
The goal is to decide whether this sentence is true.
As an example, consider the formula $\varphi(X,Y) :\equiv X^2 + Y^2 \leq 1 \land Y^2 \geq 2X^2 - 1$;
among (infinitely many) other solutions, $\varphi(0,0)$ evaluates to true, witnessing that this is a yes-instance of \ETR.
It is known that
\[
    \NP \subseteq \ER \subseteq \PSPACE
    \text{.}
\]
Here the first inclusion follows because a \problemname{SAT} instance can trivially be written as an equivalent \ETR instance.
The second inclusion is highly non-trivial and was first proven by Canny in his seminal paper~\cite{Canny1988_PSPACE}.
Interestingly, there are also oracle separation results indicating that those three complexity classes are distinct~\cite{hamm2025oracle}.

Note that the complexity of working with continuous numbers was studied in various contexts.
To avoid confusion, let us make some remarks on the underlying machine model.
The underlying machine model for \ER (over which sentences need to be decided and where reductions are performed) is the \wordRAM (or equivalently, a Turing machine) and not the \realRAM~\cite{Erickson2022_SmoothingGap} or the Blum-Shub-Smale model~\cite{Blum1989_ComputationOverTheReals}.

The complexity class \ER gains its importance by numerous important algorithmic problems that have been shown to be complete for this class in recent years.
The name \ER was introduced by Schaefer in~\cite{Schaefer2010_GeometryTopology} who also pointed out that several \NP-hardness reductions from the literature actually implied \ER-hardness.
For this reason, several important \ER-completeness results were obtained before the need for a dedicated complexity class became apparent.

Common features of \ER-complete problems are their continuous solution space and the nonlinear relations between their variables.
Important \ER-completeness results include the realizability of abstract order types~\cite{Mnev1988_UniversalityTheorem,Shor1991_Stretchability} and geometric linkages~\cite{Schaefer2013_Realizability}, as well as the recognition of geometric segment graphs~\cite{Kratochvil1994_IntersectionGraphs,Matousek2014_IntersectionGraphsER}, unit-disk graphs~\cite{Kang2012_Sphere,McDiarmid2013_DiskSegmentGraphs},    and ray intersection graphs~\cite{Cardinal2018_Intersection}.
More results appeared in the graph drawing community~\cite{Dobbins2022_areaUniversality,Erickson2019_CurveStraightening,Lubiw2022_DrawingInPolygonialRegion,Schaefer2021_FixedK}, regarding polytopes~\cite{Dobbins2019_NestedPolytopes,Richter1995_Polytopes}, the study of Nash-equilibria~\cite{Berthelsen2019_MultiPlayerNash,Bilo2016_Nash, Bilo2017_SymmetricNash,Garg2018_MultiPlayer,Schaefer2017_FixedPointsNash}, training neural networks~\cite{Abrahamsen2021_NeuralNetworks, train-fully-neural-networks}, matrix factorization~\cite{Chistikov2016_Matrix,Schaefer2018_tensorRank,Shitov2016_MatrixFactorizations,Shitov2017_PSMatrixFactorization}, or continuous constraint satisfaction problems~\cite{Miltzow2022_ContinuousCSP}.
In computational geometry, we would like to mention the art gallery problem~\cite{Abrahamsen2022_artGallery, stade2022complexity} and covering polygons with convex polygons~\cite{Abrahamsen2022_Covering}.
A complete overview can be found in a recent compendium~\cite{SchaeferCardinalMiltzow2024Compendium}.

Recently, the community started to pay more attention to higher levels of the ``real polynomial hierarchy'', which surprisingly captures some interesting algorithmic problems~\cite{Blanc2021_ESS,DCosta2021_EscapeProblem,Dobbins2022_areaUniversality,Jungeblut2022_Hausdorff,Real-Poly-Hierarchy,Burgisser2009_ExoticQuantifiers}.
And even to the first order theory of the reals as a complexity class.

%%%%%%%%%%%%%%%%%%%%%%%%%%%%%%%%%%%%%%%%%%%%%%%%%%%%%%%%%%%%%
\section{Preparation}
\label{sec:preparation}
%%%%%%%%%%%%%%%%%%%%%%%%%%%%%%%%%%%%%%%%%%%%%%%%%%%%%%%%%%%%%
We aim to construct a polygon.
This polygon should be guarded by two guards at irrational coordinates but requires three guards at rational coordinates.
We must restrict the possible coordinates the guards can be positioned. 
% Specifically, the guards should not guard the entire polygon if their coordinates are rounded to a rational number.
In this section, we will explore the tools to restrict the possible positions of the two guards within the polygon.

\begin{figure}[tbp]
    \centering
    \includegraphics[page=6]{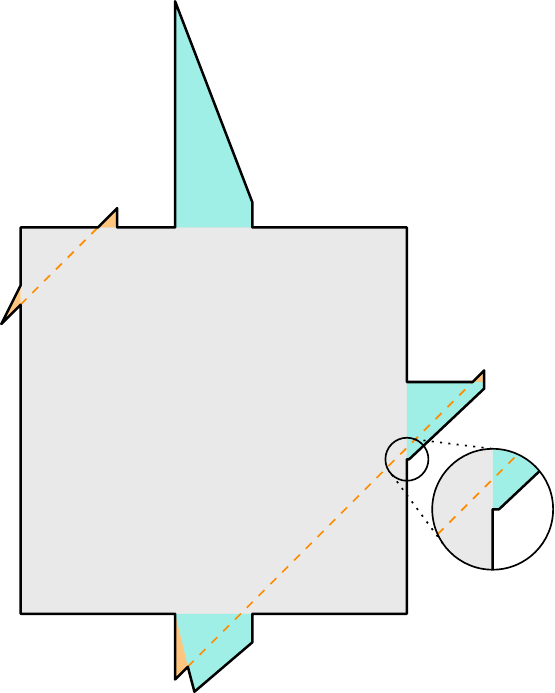}
    \caption{Our final polygon: it has a core (gray), three quadrilateral pockets (blue), and four narrow triangular pockets (yellow).}
    \label{fig:examplepolygon}
\end{figure}

\subsection{Basic Definitions}

Each guard $g$ will be able to guard some region of the polygon: 
we call this region its \emph{visibility polygon} $\vis(g)$. 
The visibility polygon includes all points for which the line segment between the guard and the point is included in the polygon $P$. 
Notably, the union of the visibility polygons of the two guards must be the art gallery. Otherwise, the art gallery is not completely guarded.

A \emph{window} is an edge of the visibility polygon $\vis(g)$ that is not part of the boundary of $P$.
We can find windows in a guard $g$'s visibility polygon, by shooting rays from $g$ to reflex vertices (the vertices of the polygon, with an interior angle larger than $\pi$).
If these rays do not leave the polygon at the reflex vertex,
% the polygon's boundary at the reflex vertex, 
a window will exist between the reflex vertex and the position where the ray does intersect the boundary of the polygon.
Let the \emph{window's end} be the intersection of the ray with an edge of the polygon.

Our final polygon consists of the core and a number of pockets, as shown in \Cref{fig:examplepolygon}. 
The \emph{core} of the polygon is the square in the center.
% of the polygon.
We will enforce that both guards are located in the core.
As a square is a convex shape, this implies that both guards will guard the core.
% 
% On the other hand, 
The \emph{pockets} are all regions outside the core. 
We will use pockets that are either quadrilateral or triangular. 
Pockets are \emph{attached} to either the core or another pocket: they have one edge that lies on the boundary of the core or on the boundary of another pocket.
Quadrilateral pockets will always be attached to the core. 
Each quadrilateral pocket has one edge that is not on the boundary of the core, nor adjacent to it.
We will call this edge the \emph{wall} of a 
quadrilateral pocket.
Similarly, triangular pockets will be attached to either the core or a quadrilateral pocket.
We will use pockets as a tool to limit the locations of the two guards.

\subsection{Guard Segments}
\begin{figure}[tbp]
    \centering
    \includegraphics[page=7]{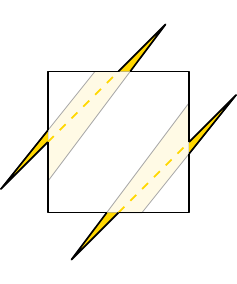}
    \caption{A small polygon that can only be guarded by two guards, because each guard segment (yellow dashed line) must contain a guard. The region where a guard could guard at least one pocket is shaded in light yellow.}
    \label{fig:guardsegment}
\end{figure}
We can force a guard to be positioned on a line segment within the polygon.
Such a line segment is called a \emph{guard segment}.
Guard segments are commonly used in the context of the art gallery problem~\cite{abrahamsen2017irrational, stade2022complexity}.
In this section, we will describe how we construct a guard segment.
Let us denote by $s$ the segment and by $\ell$ its supporting line.

To make $s$ a guard segment, we add two triangular pockets 
where $\ell$ intersects $\partial P$.
% on opposite sides of the 
% polygon.
Each of the triangular pockets has an edge on $\ell$.
Besides this one edge, the pockets lay on different sides of $\ell$.
Only a guard on the line segment between the two pockets can guard both triangular pockets at the same time.
% We can reduce the size of the line segment with a third pocket, but we will not need this in our construction.

We have two guards in our polygon and both will be on a guard segment.
If the two guard segments are not intersecting, we can enforce that there must be one guard on each of them as follows.
First, note that there are in total four triangular pockets.
Second, we make the triangular pockets sufficiently narrow.
In this way, it is impossible to guard two of the triangular pockets outside of a guard segment.
Thus at least one guard must be on each guard segment.
A simple construction with two non-intersecting guard segments is shown in \Cref{fig:guardsegment}.

% note that we can ensure no guard on one guard segment can guard a pocket of the other guard segment; any triangular pocket can be made narrow such that any guard must lie arbitrarily close to its supporting line.
% Thus, unless the guard segments intersect, we can limit the possible positions to guard any triangular pocket to not intersect either.

\subsection{Guarding Quadrilateral Pockets}
\begin{figure}[tbp]
    \centering
    \includegraphics[page=8]{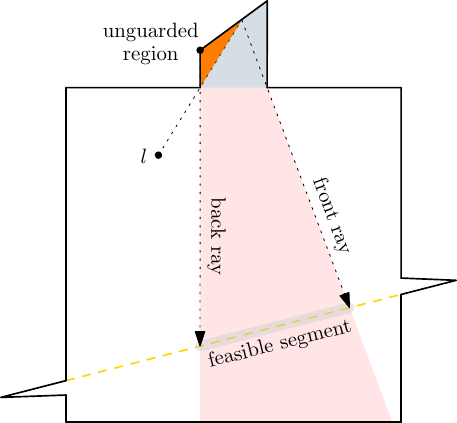}
    \caption{A polygon with guard \GuardLucas. The guard \GuardLucas defines an unguarded region in the quadrilateral pocket, a \frontRAY{} and a \backRAY{}, and a feasible segment.}
    \label{fig:simplepocket}
\end{figure}

We will now describe how given the position of guard \GuardLucas and a quadrilateral pocket $Q$ will limit the position of guard \GuardTill.
See \Cref{fig:simplepocket} for an illustration of the following description.
First, note that if \GuardLucas will not guard $Q$ completely then there will remain some unguarded region (orange) in $Q$.
The part of the guard segment of \GuardTill where the unguarded region is visible is denoted the \emph{\feasibleSegment}.
It is bounded from the \emph{\backRAY} and the \emph{\frontRAY{}}.
It is clear that \GuardTill must be on the feasible segment.

We can compute the \frontRAY{} by first computing the window end's $s$ from \GuardLucas to the wall of $Q$ and then shooting a ray from $s$ in the direction of the second reflex vertex of $Q$.

\begin{figure}[p]
    \centering
    \includegraphics[page=2,trim={0 0 0 3cm},clip]{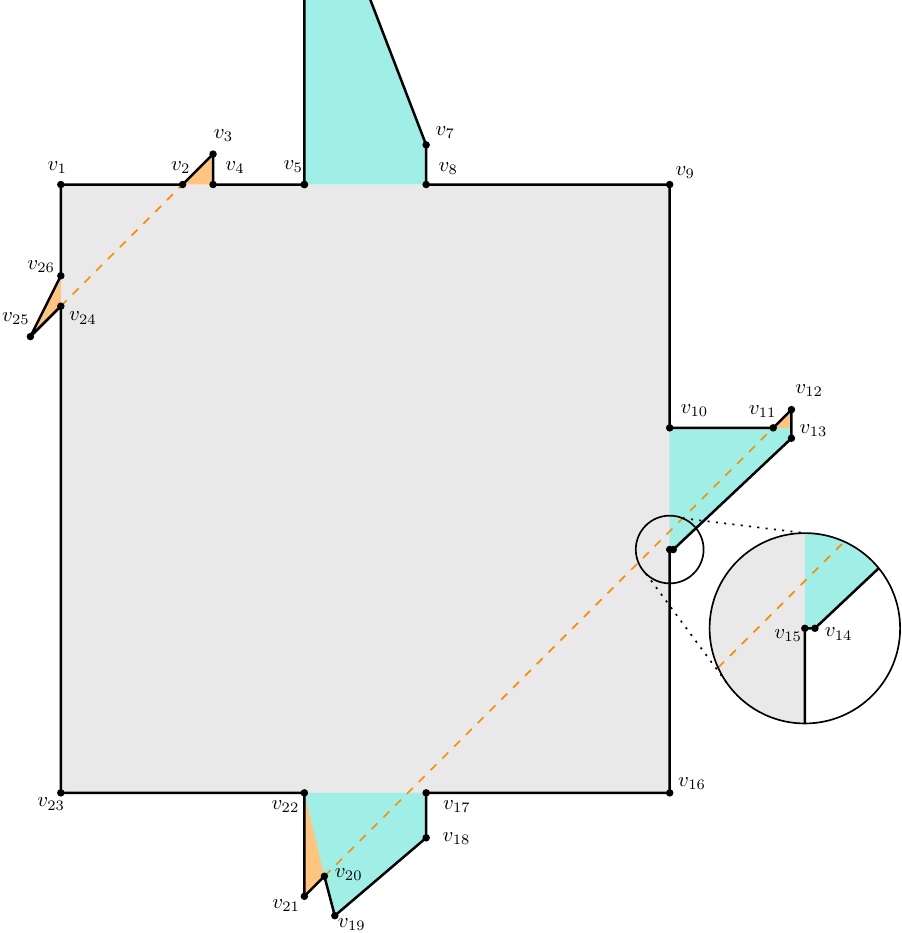}
    \caption{Our complete polygon. Note that $v_6$ is cropped. The art gallery is shaded according to the function of each region: gray is the core, yellow is the pockets used to create guard segments, and turquoise are other pockets. The yellow dashed lines represent the guard segments. The coordinates of important vertices are given.}
    \label{fig:fullpolygon1}
\end{figure}

\begin{figure}[p]
    \centering
    \includegraphics[page=4,trim={0 0 0 3cm},clip]{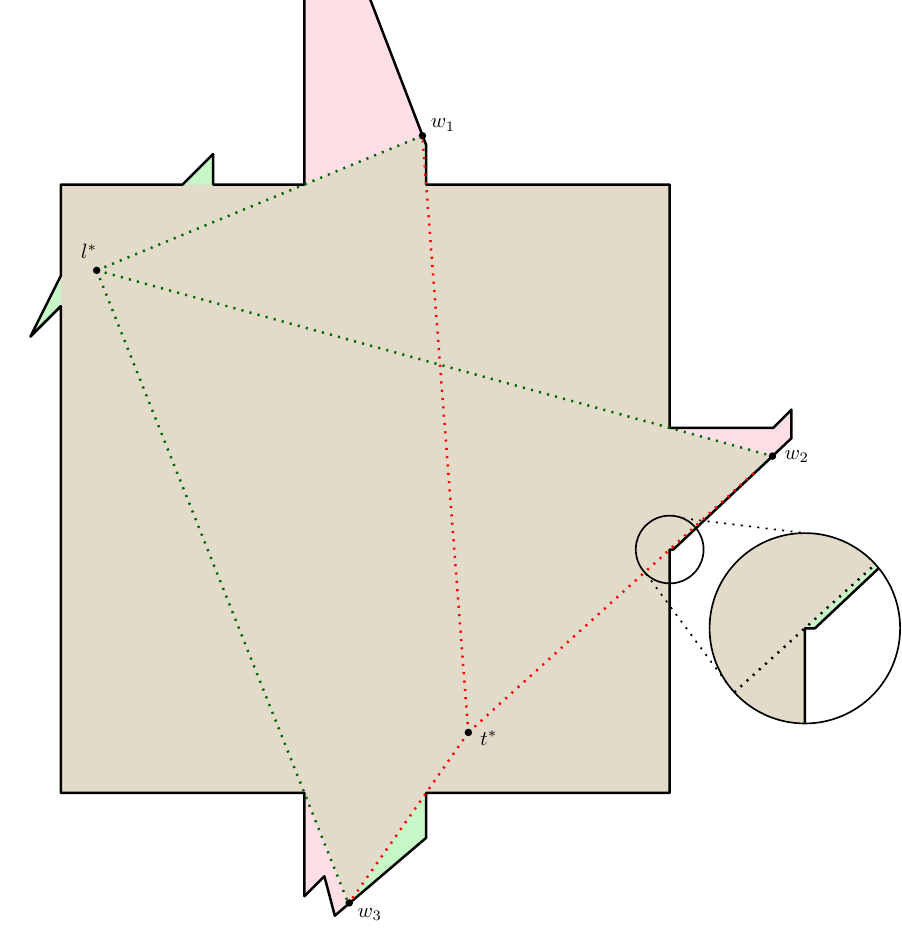}
    \caption{Our complete polygon. The optimal solution with two guards at irrational coordinates is shown. The green regions are guarded by the upper left guard; the red regions are guarded by the bottom right guard; the purple regions are guarded by both. The dashed lines are rays shot from the guards through reflex vertices. For each pocket, these windows meet at a point on the art gallery's wall, of which the coordinates are also given.}
    \label{fig:fullpolygon2}
\end{figure}
\begin{table}[H]
\centering
    \caption{Coordinates of the vertices of the polygon ($v_1, \dots, v_{26}$), the guards (\GuardLucas and \GuardTill), and the window's ends ($w_1, w_2, w_3$).}
\begin{tabular}{|ll|ll|ll|}
\hline
$v_1$    & $(0, 10)$                   & $v_{12}$ & \textbf{$(12, 6.3)$}           & \textbf{$v_{22}$} & $(4, 0)$                                                                                         \\
$v_2$    & $(2, 10)$                   & $v_{13}$ & $(12, \frac{624712}{107015})$  & $v_{23}$          & $(0, 0)$                                                                                         \\
$v_3$    & $(2.5, 10.5)$               & $v_{14}$ & $(\frac{1016072}{101007}, 4)$  & $v_{24}$          & $(0, 8)$                                                                                         \\
$v_4$    & $(2.5, 10)$                 & $v_{15}$ & $(10, 4)$                      & $v_{25}$          & $(-0.5, 7.5)$                                                                                    \\
$v_5$    & $(4, 10)$                   & $v_{16}$ & $(10, 0)$                      & $v_{26}$          & $(0, 8.3)$                                                                                       \\
$v_6$    & $(4, \frac{465522}{29357})$ & $v_{17}$ & $(6, 0)$                       & $\GuardLucas^*$   & $(3.7-2.2\cdot\sqrt{2}, 11.7-2.2\cdot\sqrt{2})$                                                  \\
$v_7$    & $(6, \frac{312388}{29357})$ & $v_{18}$ & $(6, \frac{-25442}{34407})$    & $\GuardTill^*$    & $(7.4-0.5\cdot\sqrt{2}, 1.7-0.5\cdot\sqrt{2})$                                                   \\
$v_8$    & $(6, 10)$                   & $v_{19}$ & $(4.5, \frac{-138913}{68814})$ & $w_1$             & $(\frac{293570\cdot\sqrt{2}+8052346}{1425913}, \frac{-765670\cdot\sqrt{2}+16485384}{1425913})$   \\
$v_9$    & $(10, 10)$                  & $v_{20}$ & $(4.33, -1.37)$                & $w_2$             & $(\frac{1071750\cdot\sqrt{2}+29733818}{2673483}, \frac{1010070\cdot\sqrt{2}+13370606}{2673483})$ \\
$v_{10}$ & $(10, 6)$                   & $v_{21}$ & $(4, -1.7)$                    & $w_3$             & $(\frac{344070\cdot\sqrt{2}+3108526}{760803}, \frac{293430\cdot\sqrt{2}+1804526}{760803})$       \\
$v_{11}$ & $(11.4, 6)$                 &          &                                &                   &                                                                                                  \\ \hline
\end{tabular}

    \label{tab:my_label}
\end{table}

%%%%%%%%%%%%%%%%%%%%%%%%%%%%%%%%%%%%%%%%%%
\section{Complete Polygon}
\label{sec:Polygon}
%%%%%%%%%%%%%%%%%%%%%%%%%%%%%%%%%%%%%%%%%%
In this section, we will present our complete polygon: a polygon that can be guarded by two guards if and only if both guards are situated at irrational points. 
% We prove that two rational guards cannot guard the polygon.

\subsection{The Polygon}
As we described in \Cref{sec:preparation} and displayed in \Cref{fig:fullpolygon1}, the polygon consists of a core and some pockets.
The polygon has four triangular pockets defining two guard segments.
The two guard segments lie on the lines $y=x+8$ and $y=x-5.7$.
Furthermore, the polygon has three quadrilateral pockets.
In \Cref{tab:my_label}, the coordinates of the vertices of the polygon, the coordinates of the two guards, and the coordinates of the window's ends are given.

The walls of the three quadrilateral pockets have the supporting lines:
\begin{enumerate}
    \item Top pocket: \begin{math}
        y=\frac{-76567\cdot x+771790}{29357}.
    \end{math}
    \item Right pocket: \begin{math}
        y=\frac{101007\cdot x-587372}{107175}.
    \end{math}
    \item Bottom pocket: \begin{math}
        y=\frac{29343\cdot x-201500}{34407}.
    \end{math}
\end{enumerate}

\subsection{Proof of \Cref{thm:main}}
\begin{proof}
We prove that our polygon can be guarded with two irrational guards, but cannot be guarded with two rational guards. 
We state that the polygon can be guarded by two guards placed at 
\begin{displaymath}
    \GuardLucas^\star= (3.7-2.2\cdot\sqrt{2}, 11.7-2.2\cdot\sqrt{2})
\end{displaymath} and \begin{displaymath}
    \GuardTill^\star = (7.4-0.5\cdot\sqrt{2}, 1.7-0.5\cdot\sqrt{2}).
\end{displaymath} 
\Cref{fig:fullpolygon2} displays the visibility polygons of these two guards. 
It can be checked using simple calculations that the two visibility polygons cover the complete polygon.
First, both guard segments will contain a guard. 
Furthermore, the window's ends are in the same location (so no unseen region between them) and both vertices on the wall are guarded.

Now, we prove that no two rational guards can guard our polygon. 
Specifically, we show that no two guards, except the ones mentioned, will guard the entire polygon. 
Clearly, both guard segments must contain a guard. 
Let \GuardLucas be on the guard segment in the top left of the polygon, and \GuardTill be on the guard segment in the bottom right of the polygon. 
Given the position of \GuardLucas, we calculate where \GuardTill can guard all regions not guarded by \GuardLucas.
For each of the three pockets, we will bound the position of \GuardTill given the position of \GuardLucas. 
To represent their positions, we will use their x-coordinates. 
As both \GuardLucas and \GuardTill lie on a non-vertical guard segment, their x-coordinates will uniquely describe their position.

First, we calculate which part of the pocket \GuardLucas guards and which part it does not. 
Due to the guard segment of \GuardLucas, we will have exactly one window along with its corresponding window's end. 
As described in \Cref{sec:preparation}, we can use this construction to determine the region where \GuardTill can guard the region of the pocket \GuardLucas does not cover.
Notably, \GuardTill will always be on the correct side of the \backRAY.
Indeed, the entire guard segment of \GuardTill lies on one side of the \backRAY. 
As such, we will bound the \feasibleSegment by ensuring \GuardTill lies on the correct side of the \frontRAY.
We calculate the intersection of this ray and the guard segment.
Then, the x-coordinate of \GuardTill may either not be smaller than, or be greater than the x-coordinate of the intersection between the \frontRAY and the guard segment.
\begin{figure}[tbp]
    \centering
    \includegraphics[page=10]{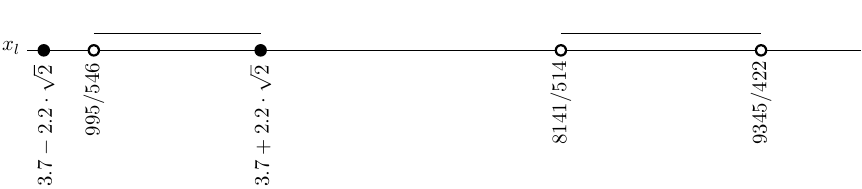}
    \caption{The solution to the system of equations (\Cref{eq:1}, \Cref{eq:2}, \Cref{eq:3}). Here, $\bullet$ denotes a closed interval, while $\circ$ denotes an open interval.}
    \label{fig:proof}
\end{figure}
It depends on the pocket whether the x-coordinate of \GuardTill can not be smaller or greater than the x-coordinate of the intersection.
Guard \GuardTill must lie on the same side of the \frontRAY as the unguarded region of the pocket.
As can be verified in \Cref{fig:fullpolygon2}, the x-coordinate of \GuardTill interacts with the intersections we find for the pockets in the following way:
\begin{enumerate}
    \item Top pocket: the x-coordinate of \GuardTill must be smaller or equal to the intersection.
    \item Right pocket: the x-coordinate of \GuardTill must be greater or equal to the intersection.
    \item Bottom pocket: the x-coordinate of \GuardTill must be smaller or equal to the intersection.
\end{enumerate}
It is important to note that the x-coordinate of \GuardTill does not lie on the same side of all three intersections.
If it did lie on the same side of all three, then the position of \GuardTill could trivially be at any coordinate greater or smaller than all three intersections.

We can use the x-coordinate of \GuardLucas to determine its position. 
So, we use \GuardLucas's x-coordinate ($x_\GuardLucas$) to calculate inequalities that limit the x-coordinate of \GuardTill ($x_\GuardTill$):
\begin{align}
    x_\GuardTill\leq \frac{42734\cdot x_\GuardLucas-70239}{5460\cdot x_\GuardLucas-9950} \label{eq:1} \\
    x_\GuardTill\geq \frac{11928\cdot x_\GuardLucas-269330}{2570\cdot x_\GuardLucas-40705} \label{eq:2} \\
    x_\GuardTill\leq \frac{13538\cdot x_\GuardLucas-616793}{4220\cdot x_\GuardLucas-93450} \label{eq:3}
\end{align}
We will use \Cref{eq:1}, \Cref{eq:2}, \Cref{eq:3} as a system of equations.
A solution to the system of equations will have a corresponding pair of guards.
We use an algebraic computer program to calculate the solution to this system of equations.
\Cref{fig:proof} shows the values for $x_\GuardLucas$ for which the system of equations has a valid solution.
Then, any value for $x_\GuardTill$ is chosen between the bounds imposed by $x_\GuardLucas$.
However, not all solutions for $x_\GuardLucas$ correspond to valid positions for guard \GuardLucas.
Specifically, notice in \Cref{fig:proof} that the only possible value for $x_\GuardLucas < \frac{10539}{9974} < \frac{995}{546}$ is irrational.
We will argue that $x_\GuardLucas$ must be smaller than $\frac{10539}{9974}$.
For the following description, we refer to \Cref{fig:Left-Bounding-Guard}.

\begin{figure}
    \centering
    \includegraphics[page=5]{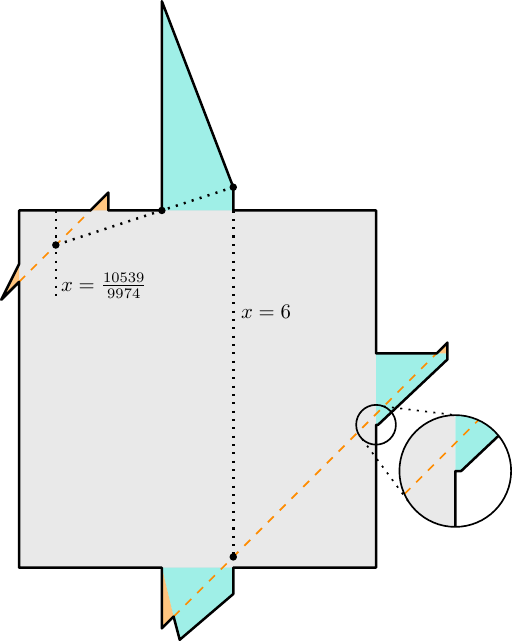}
    \caption{The guard \GuardLucas must be to the left of the line  $x = \frac{10539}{9974}$.}
    \label{fig:Left-Bounding-Guard}
\end{figure}

Suppose for the purpose of contradiction that there is a valid guard placement with $x_\GuardLucas\geq\frac{10539}{9974}$.
In this case, guard \GuardLucas fails to guard any part of the wall of the top pocket.
Guard $x_\GuardTill$ must be less than or equal to $6$ to guard the entire wall.
However, for $x_\GuardTill$ less than or equal to $6$, guard \GuardTill fails to guard the wall of the right pocket. % Would only intersect at x=13.2094
Guard \GuardLucas can never guard the entire wall of the right pocket.
This gives a contradiction and implies that
 $x_\GuardLucas < \frac{10539}{9974}$.

As such, we can limit the possible locations for guard \GuardLucas as $x_\GuardLucas < \frac{10539}{9974}$. 
Evidently, in this range, the only valid x-coordinate for $x_\GuardLucas$ is $3.7-2.2\cdot\sqrt{2}$, see \Cref{fig:proof}. 
For this x-coordinate of $x_\GuardLucas$, the only possible position for \GuardTill is at $x_\GuardTill = 7.4-0.5\cdot\sqrt{2}$.
Finally, this shows that the only possible configuration of two guards in this polygon is at $(3.7-2.2\cdot\sqrt{2}, 11.7-2.2\cdot\sqrt{2})$, and at $(7.4-0.5\cdot\sqrt{2}, 1.7-0.5\cdot\sqrt{2})$: both guards must be at irrational coordinates.
\end{proof}

%%%%%%%%%%%%%%%%%%%%%%%%%%%%%%%%%%%%%%%%%%%
\section{Challenges}
\label{sec:Challenges}
%%%%%%%%%%%%%%%%%%%%%%%%%%%%%%%%%%%%%%%%%%%
We encountered new challenges while searching for our polygon, compared to Abrahamsen, Adamaszek, and Miltzow~\cite{abrahamsen2017irrational}'s polygon that requires three irrational guards.

\begin{figure}[tbp]
    \centering
    \includegraphics[page=9]{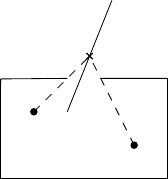}
    \hspace{2cm}
    \includegraphics[page=11]{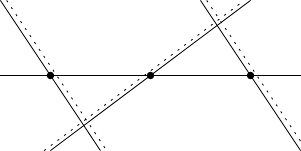}
    \caption{Left: Two guards and two reflex vertices, where the corresponding wall leads to an invalid polygon. Right: Three \frontRAY{}s intersecting the guard segment. 
    Guard \GuardTill must be on the dotted side of each ray. 
    If the intersection points were reversed, there does not exist a valid solution either.}
    \label{fig:properites}
\end{figure}

\paragraph{Irrational Lines.}
To describe how we found the polygon described in the previous section, we introduce the concept of a rational and an irrational line.
Give a line $\ell = \{mx + n : x\in \R \}$, we say that $\ell$ is \emph{rational} if and only if $m,n\in \Q$, and \emph{irrational} otherwise.
Note that a rational line might contain arbitrarily many irrational points. (For example, the line $y=0$ contains all the irrational points of the form $(a,0)$, with $a$ being irrational.)
Now, if we have two distinct rational points, then they span a rational line.
And by point-line duality, if we have two rational lines, then they intersect at a rational point. 
This implies that any irrational point is contained in at most one rational line.
And conversely, any irrational line contains at most one rational point.
(For instance, the point $(\pi,\pi)$ is on the rational line $x=y$.)

This is relevant to us, as our guards are irrational points.
Our guard segments are rational lines, as they are defined by triangular pockets. 
And the triangular pockets consist of rational vertices.
Specifically, any window's end must be irrational, for the following reason.
The line $\ell$ defined by the guard and the reflex vertex must be irrational, as it is not the guard segment.
Furthermore, $\ell$ already contains a rational point, namely the reflex vertex. 
Thus, the window's end must be irrational as well.

\paragraph{Constructing the Polygon.}
To start, like in the construction with three guards, we parameterize the position of the (irrational) positions of the two guards and the reflex vertices before creating the polygon. 
In other words, $g = (g_1\cdot\sqrt{2} + g_2, g_3\cdot\sqrt{2} + g_4)$, with $g_i \in \Q$.
To define the guard segments, we compute the unique rational line through each guard.
Then, we define the quadrilateral pockets. 
To do this, we note that the window's ends of the two guards must meet on each pocket's wall. 
For a pocket, its window's end comes from two rays shot from the two guards through the reflex vertices of the pockets.
Let $w$ be the point where the two rays intersect.
Recall that $w$ will always be an irrational point.
One unique rational line will contain $w$, so this must be the supporting line of the wall.
Now, changing the coordinates of the two parameterized guards and the reflex vertices will change the position of the guard segments and the walls of all the quadrilateral pockets.
Note that not all walls and guard segments lead to a valid polygon, or to a polygon at all.
We have to check a few properties that we will describe next.

\paragraph{Properties.}
Here, we list some properties that we need to ensure.
We start with the properties of the guard segments.
As already mentioned, the two guard segments should not intersect.
Furthermore, if the guard segments go across the whole polygon,
it might be possible to guard all quadrilateral pockets with a single guard.
Thirdly, if a guard segment is aligned such that a larger region of each pocket is observed when the guard is moved in one direction, then the interaction with the other guard is not meaningful.
As such, many positions of guards cause invalid guard segments.

Now, let's move to the properties of the quadrilateral pockets.
For each pocket, we ensured (by construction) that the window's ends of the two guards are at the exact same point.
% Otherwise, the two guards could still guard the pockets if their coordinates were rounded.
To create the walls of the pockets, we cast a ray from each guard to each pocket, to illustrate where the window in its visibility polygon would be.
For each pocket, we found the (irrational) intersection of these rays of the two guards.
Then, we define the wall of the pocket as a line segment on the rational line through the intersection.

This construction of the walls also poses a problem.
As mentioned above, there only exists a single rational line through an irrational point.
% The two rays will both be irrational, as will  their intersection.
Thus, there will be only one possible wall for the given guards and reflex vertices.
However, not all walls will be valid.
% : the two guards need to be on the same side of the rational line.
An example of an invalid wall is shown in the left picture in \Cref{fig:properites}.

\begin{figure}[tbp]
    \centering
    \includegraphics{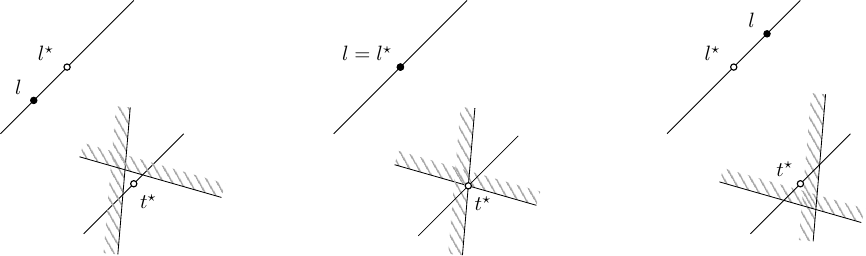}
    \caption{We draw the two guard segments without the surrounding polygon.
    On the top, we move the guard \GuardLucas from left to right.
    On the bottom, we see the corresponding feasible segment for guard \GuardTill.
    Recall that we denote by $\GuardLucas^{\star}$ and 
    $\GuardTill^\star$ the coordinates of the predetermined guard positions.}
    \label{fig:Bad-frontRAYs}
\end{figure}

Even a valid polygon does not guarantee that the construction will work.
For the following description, refer to \Cref{fig:Bad-frontRAYs}.
Our construction guarantees that the feasible segment for guard \GuardTill is limited to a single point, given the predetermined position of $\GuardLucas = \GuardLucas^*$.
(Recall that the feasible segment is an interval on the guard segment of \GuardTill).
However, it does not guarantee that there does not exist a feasible segment for \GuardTill when \GuardLucas is arbitrarily close to its predetermined position.
Recall that we determine the feasible segment for \GuardTill based on the \frontRAY{}s.
Guard \GuardTill must lie on the correct side of all three of these \frontRAY{}s.
So, when $x_\GuardLucas$ is less than its predetermined value, there must not exist an interval on the guard segment of \GuardTill that lies on the correct side of all these rays.
(Recall from \Cref{sec:Polygon} that we note $\GuardLucas = (x_\GuardLucas,y_\GuardLucas)$.)
Otherwise, there must exist a rational point for \GuardTill on the interval.
For this rational point of \GuardTill, we can find a corresponding rational point for \GuardLucas.
However, when $x_\GuardLucas$ is greater than its predetermined position, the order of the intersection of these rays with the guard segment will reverse.
So there must not exist a segment on the guard segment that lies on the correct side of all rays when the intersection points are reversed, either.

To prevent this, we arrange the \frontRAY{}s' intersection pattern as shown in the right picture of \Cref{fig:properites}.
This implies that the middle intersection requires the guard to be on one side, and the first and the last intersection require the guard to be on the other side.

\paragraph{ETR-Formula}
The Art Gallery problem is \ER-complete, which implies that it can be encoded into an \ETR-formula.
In this section, we will describe how to encode our instance into a logical formula of different types.

\paragraph{Using Universal Quantifiers.}
We first we describe a formula that also allows universal quantifiers of modest size.
% Afterward, we explain how the universal quantifiers are eliminated and that the approximate length of the resulting \ETR-formula is vast.

Allowing for universal quantifiers, we get the following simple formula~\cite{EfratH06}:
\[ \exists x_1, y_1, x_2, y_2 \forall p_x, p_y : \textrm{INSIDE-POLYGON}(p_x, p_y) \Rightarrow (\textrm{SEES}(x_1, y_1, p_x, p_y) \lor \textrm{SEES}(x_2, y_2, p_x, p_y)). \]
Here, $\textrm{INSIDE-POLYGON}(a,b)$ determines if the point $(a,b)$ is inside the polygon and 
$\textrm{SEES(a,b,x,y)}$ verifies whether the line segment between the point $(a,b)$ and the point $(x,y)$ intersects the polygon.
To implement $\textrm{INSIDE-POLYGON}(a,b)$, we find a triangulation of the polygon and determine whether $(a, b)$ is contained in any one of the triangles of the triangulation.
A short calculation shows that this formula would require more than $4000$ characters, if we assume that every variable is represented by one character only.

\paragraph{Geometric Construction.}
However, an \ETR-formula does not allow the universal quantifiers used above.
Luckily, Abrahamsen, Adamaszek, and Miltzow found a method to describe an instance of the Art Gallery problem without universal quantifiers.
They find a set of points, of which some depend on the coordinates of the guards, such that the every possible point in the polygon is guarded if and only if every point in this set is guarded.
We find this set of points as follows:
\begin{enumerate}
    \item Find a set of lines that contains the lines defining the line segments of the polygon, and every line containing both a guard and a vertex of the polygon.
    \item Find all intersections of two lines.
    \item For every triplet of intersections, find the centroid of the triangle with the three intersections as its corners.
\end{enumerate}

Our instance of the Art Gallery problem has $26$ vertices and $2$ guards, so we procedurally generate $78$ lines.
Then, there are $78^2=6084$ intersections of lines and $6084^3>10^{11}$ points in our final set.
(Indeed, we could prune some lines, due to some lines defining line segments of the polygon are equivalent.
However, the length of the \ETR-formula will still explode very quickly.)
The final ETR-formula can generate all those points and checks for each of them that if they are inside the polygon that one of the guards sees the point. 
As the number of points is huge, so will be the size of the formulae.

\paragraph{Quantifier Elimination.}
The art gallery problem has a much simpler formulation when we allow universal quantifiers as seen above.
So, another approach to find an \ETR-formula of reasonable length for our instance of the art gallery problem is to use the formulation with universal quantifiers  and perform quantifier elimination using some known algorithm.
However, despite the formulation only using four existential quantifiers and two universal quantifiers we aim to eliminate, the resulting formula is seemingly infeasibly large as we will see.
To illustrate this, we make some ``back-of-the-envelope'' calculation.
There are several algorithms readily available in the literature;
we use the algorithm from Basu et al.~\cite{Basu2006_RealAlgebraicGeometry}
as this is one of the most modern treatments of algorithms in real algebraic geometry.
The results could vary for other algorithms.
To give an upper bound on the running time, we denote
 with $k$ the numbers of existential quantifiers, with $l$ the number of universal quantifiers that we want to eliminate, with $s$ the number of polynomials, and $d$ denotes the degree of the polynomials.
Clearly, $k=4$, and $l=2$. 
Furthermore, as the formula deals with lines and determinants only, $d=2$.
The upper bound is given by 
\[s^{(k+1)\cdot(l+1)}\cdot d^{\mathcal{O}(k\cdot l)} \cdot s^{k+1}\cdot d^{O(k)} \cdot d^{\mathcal{O}(k)}.\]
We do not know the precise implicit constant $c$ in the big-Oh notation.
We use the modest estimate of $c=2$. 
However, $s$ is much larger, which we estimate next.
The predicate $\textrm{INSIDE-POLYGON}$ requires a triangulation of the polygon.
Standard techniques would find a triangulation with $24$ triangles, due to the polygon having $26$ vertices.
However, due to various line segments being collinear, we can find a triangulation with only $12$ triangles.
For each triangle, $\textrm{INSIDE-POLYGON}$ checks the determinant of the point for each of the three lines.
As such, $\textrm{INSIDE-POLYGON}$ requires $3\cdot12=36$ polynomials.
Then, $\textrm{SEES}$ determines whether the line segment between two points intersects any line segment defining the polygon.
To check whether the two line segments intersect, the order type of four triplets of points need to be checked.
As there are $26$ line segments, each requiring four checks, we require $26\cdot 4=104$ polynomials for $\textrm{SEES}$.
So, we get that $s=140$.
Using those numbers the algorithm returns an \ETR-formula of length at most 
\[\approx 140^{75}\cdot 2^{12} \approx 10^{164}.\]
Note that this is vastly more than the number of atoms of the universe.
There might be some observations that allow us to use fewer polynomials, and maybe there are faster quantifier eliminations algorithms available.
Yet, this sloppy calculation gives a clear idea on how tricky it is to generate a concise \ETR-formula in this naive way.
Furthermore, it also displays the inefficiency of quantifier elimination on even simple instances; the resulting \ETR-formula is several orders of magnitude larger than the more complex formula from Abrahamsen, Adamaszek, and Miltzow.

On a brighter note, we note that, if we use some basic observations about the polygon (such as the guards being on the line segments), the formulae \cref{eq:1}, \cref{eq:2}, and \cref{eq:3} together with range constraints suffice in determining whether there exists a solution with just two guards in our polygon. 
This system can easily be solved by a human.

\section{Conclusion.}
In comparison to the polygon construction with three guards~\cite{abrahamsen2017irrational}, our polygon has fewer parameters that we can adjust, as we have one guard less.
As everything depends on those few parameters, it was difficult to find a configuration that satisfies all the desired properties simultaneously.
Furthermore, we need to check some additional properties that did not play a role in the previous construction, as the middle guard was surrounded by the other guards from two different sides.
At last, we could not avoid the supporting line of the guard segment intersecting two quadrilateral pockets.

\acknowledgements
We would like to thank Thekla Hamm and Ivan Bliznets for their helpful comments on the presentation.

%%%%%%%%%%%%%%%%%%%%%%%%%%%%%%%%%%%%%%%%%%%%%%%%%%%%
% \newpage
\bibliographystyle{plainnat}
\bibliography{ETR, ArtGallery}

\end{document}